\documentstyle[aaspp4]{article}

\received{}
\accepted{}
\journalid{VOL}{JOURNAL DATE}
\articleid{START PAGE}{END PAGE}
\paperid{MANUSCRIPT ID}

\lefthead{Yoshida \& Lee}
\righthead{$R$-modes of neutron stars with a solid crust}

\begin{document}


\title{$R$-modes of neutron stars with a solid crust}
\author{Shijun Yoshida\altaffilmark{1} and Umin Lee}
\affil{Astronomical Institute, Graduate School of Science, 
Tohoku University, Sendai 980-8578, 
Japan \\ yoshida@astr.tohoku.ac.jp, lee@astr.tohoku.ac.jp}

\altaffiltext{1}{Research Fellow of the Japan Society for 
the Promotion of Science.}

\begin{abstract}

We investigate the properties of $r$-mode oscillations of a slowly rotating
neutron star with a solid crust, by taking account of the effects of
the Coriolis force.
For the modal analysis we employ three-component neutron star models 
that are composed of a fluid core, a solid crust and a surface fluid ocean.
For the three-component models, we find that
there exist two kinds of $r$-modes, that is, those confined 
in the surface fluid ocean and those confined in the fluid core, which
are most important for the $r$-mode instability.
The $r$-modes do not have any appreciable amplitudes in the solid crust if 
rotation rate of the star is sufficiently small.
We find that the core $r$-modes are strongly affected by mode 
coupling with the crustal torsional (toroidal) modes and lose their simple properties of
the eigenfunction and eigenfrequency
as functions of the angular rotation velocity $\Omega$. 
This indicates that the extrapolation formula, which is obtained
in the limit of $\Omega\rightarrow 0$, cannot be used to
examine the $r$-mode instability of rapidly rotating neutron stars 
with a solid crust 
unless the effects of mode coupling with the crustal torsional modes 
are correctly taken into account.

\end{abstract}

\keywords{instabilities --- stars: neutron --- 
stars: oscillations --- stars: rotation}


\section{Introduction}

Since the discovery of the gravitational radiation driven instability of
the $r$-modes by Andersson (1998) and Friedman \& Morsink (1998),
a large number of studies on the properties of $r$-modes and inertial modes 
of rotating stars have been done to prove their possible 
importance in astrophysics (for a recent review see, e.g., Friedman \& Lockitch 1999). 
Although early investigations of the $r$-mode instability were mainly applied to
young hot neutron stars (e.g., Lindblom, Owen, \& Morsink 1998; Owen et al. 1998), 
recently some authors have also discussed possible roles of
the $r$-mode instability in old and cool neutron stars 
with a solid crust and a magnetic field, as those found in low mass X--ray binaries 
(LMXBs) (e.g., Andersson, Kokkotas, \& Stergioulas 1999;
Bildsten \& Ushomirsky 2000; Rezzolla, Lamb, \& Shapiro 2000).
For example, Bildsten \& Ushomirsky (2000) have suggested that the $r$-mode instability
can be largely weakened by the effects of viscous damping 
in the boundary layer at the interface between the solid crust and the fluid core.
Rezzolla et al. (2000) have also suggested that amplitude of $r$-mode oscillations tends to 
reduce due to the coupling to the magnetic field of a star.

It is well known that neutron stars with a solid crust show a rich spectrum of 
nonradial oscillation modes, for example, nonradial modes associated with the solid crust 
and those associated with the fluid core and the surface fluid ocean 
(McDermott et al. 1988).
If the effects of rotation are taken into account, it is quite common to find 
avoided crossings, as functions of $\Omega$, between various oscillation modes
of the neutron star model with a solid crust (Lee \& Strohmayer 1996).
This is also the case for $r$-modes.
When the angular rotation frequency $\Omega$ of the star is small, the $r$-mode frequency
observed in the corotating frame may be given by
\begin{equation}
\omega_r(\Omega)\approx {2m\Omega\over l'(l'+1)} \, , 
\label{om-r}
\end{equation}
where $l'$ and $m$ are the indices of the spherical harmonic function representing
the dominant toroidal component of the displacement vector.
On the other hand, the crustal toroidal mode frequency may be given by
(e.g., Strohmayer 1991)
\begin{equation}
\omega_t(\Omega)\approx \omega_t(0)+{m\Omega\over l'(l'+1)} \, ,
\label{om-t}
\end{equation}
where $\omega_t(0)$ is the oscillation frequency of the mode at $\Omega=0$.
The frequencies of the two modes get close to each other at
\begin{equation}
\Omega_{cross}\approx \frac{l'(l'+1)}{m} \, \omega_{t}(0) \, . 
\label{om-cross}
\end{equation}
As shown by McDermott et al. (1988) and Lee \& Strohmayer (1996),
the oscillation frequency of the crustal toroidal mode that has no radial nodes of 
the eigenfunction is of order of $\omega_t(0)/\sqrt{GM/R^3}\sim 10^{-2}$
for $l'=2$, and hence we have 
$\Omega_{cross}/\sqrt{GM/R^3}\sim10^{-2}\times (l'+1)$
for the modes with $l'=m$, where $M$ and $R$ are the mass and radius of the 
neutron star, and $G$ is the gravitational constant.
This suggests that the mode crossing can happen between 
the two modes at sufficiently small rotation rates. 
Almost all previous discussions on the
$r$-mode instability, however, did not take account of the effects of mode crossing
between the two different kinds of modes.\footnote{After our submission of this 
paper, a paper by Levin \& 
Ushomirsky (2000) concerning a similar problem appeared in a preprint server 
``astro-ph''. They treated $r$-modes in a constant density neutron star with a constant 
shear modulus crust.} 
In fact, the discussions have been based on the assumptions
that the modal property of the $r$-modes is almost independent of the rotation frequency 
$\Omega$ and hence that an extrapolation formula, obtained 
in the limit of $\Omega\rightarrow 0$,
is applicable to rapidly rotating neutron stars with $\Omega/\sqrt{GM/R^3}\sim 1$.
Obviously these two assumptions are not necessarily correct for the
$r$-modes that experience mode coupling with other kinds of oscillation modes.

In this paper, we examine the $r$-mode instability of neutron stars that have
a solid crust, by considering the effects of the mode crossing between the $r$-modes and
the crustal toroidal modes.
The plan of this paper is as follows. In \S 2, we show the modal properties of
the $r$-modes of neutron stars with a solid crust.  In \S 3, we discuss 
the dissipation timescales of the $r$-modes. \S 4 is for discussions and
conclusions.


\section{$r$-Modes of Neutron Stars with a Solid Crust}

Neutron star models we use in this paper are the same as those
used in the modal analysis by McDermott et al. (1988).
The models are taken from the evolutionary sequences 
for cooling neutron stars calculated by Richardson 
et al. (1982), where the envelope structure is constructed 
by following Gudmundsson, Pethick \& Epstein (1983). 
These models are composed of a fluid core, a solid crust and a surface 
fluid ocean, and the interior temperature is finite and is not 
constant as a function of the radial distance $r$. 
The models are not isentropic and
the Schwarzschild discriminant $A$ has finite values in the interior 
(see, e.g., Yoshida \& Lee 2000b).

The method of calculation of nonradial oscillations of rotating neutron stars 
with a solid crust is the same as that used by Lee \& Strohmayer (1996), in which
the Cowling approximation is employed.
The terms due to the Coriolis force are included in the perturbation equations
but no effects of the centrifugal force are considered.
This approximation may be justified for low frequency modes satisfying
$\left|2\Omega/\omega\right|\ge 1$ and $\Omega^2/(GM/R^3)\ll1$, where $\omega$ is the
oscillation frequency observed in the corotating frame of the star
(Unno et al. 1989).
The eigenfunctions are expanded in terms of spherical harmonic
functions $Y^m_l(\theta,\varphi)$ with different values of $l$ for a given $m$.
Here spherical polar coordinates $(r,\theta,\varphi)$ are used.   
For example, the Lagrangian displacement vector, $\xi_i$ is expanded as
\begin{equation}
\xi_r = r \sum_{l\geq\vert m \vert}^{\infty}  \, S_l(r)  
Y_l^m (\theta,\varphi)  e^{i \sigma t} \,  , 
\label{xi-r}
\end{equation}
\begin{equation}
\xi_\theta = r \sum_{l,l'\geq\vert m \vert}^{\infty} 
\left\{ H_l (r) {\partial {Y_l^m}\over\partial\theta}  
+ T_{l'} (r) \frac{1}{\sin \theta} \, 
{\partial{Y_{l'}^m}\over\partial\varphi} \right\} e^{i \sigma t} \, ,
\label{xi-th}
\end{equation}
\begin{equation}
\xi_\varphi =  
r \sum_{l,l'\geq\vert m \vert}^{\infty} \left\{ H_l (r) {1\over\sin\theta}
{\partial {Y_l^m}\over\partial\varphi}
        - T_{l'} (r) {\partial {Y_{l'}^m}\over\partial\theta}
        \right\} e^{i \sigma t} \, , 
\label{xi-ph}
\end{equation}
where $l=|m|+2k$ and $l'=l+1$ for even modes and $l=|m|+2k+1$ and $l'=l-1$
for odd modes where $k=0,~1,~2~\cdots$
(Lee \& Strohmayer 1996, see also Lee \& Saio 1986).
The symbol $\sigma$ denotes the oscillation frequency observed in an inertial frame,
and we have $\omega=\sigma+m\Omega$.
In this paper, to obtain a good convergence of the
eigenfunctions and eigenfrequencies at a given $\Omega$,
we keep an enough number of terms
in the series expansion of the eigenfunctions.

We computed frequency spectra of $r$-modes, inertial modes, and crustal toroidal modes
for the three-component neutron star models called
NS05T7, NS05T8, and NS13T8 (see, McDermott et al. 1988). 
The obtained mode spectra for the three models are qualitatively the same.
In this paper, we therefore show the mode spectrum for the model 
NS13T8, because this is the most compact model among the three. 
The mass $M$ and the radius $R$ of the model NS13T8 are $M=1.326 \ M_{\sun}$ and 
$R=7.853 \ \rm{km}$, respectively, and
the central temperature is $T_c=1.05 \times 10^8 \ \rm{K}$.

In Figure 1, scaled eigenfrequencies $\kappa\equiv \omega/\Omega$ of $r$-modes, inertial modes,
and a crustal toroidal mode of the model NS13T8 
are given as functions of $\bar\Omega\equiv\Omega/\sqrt{GM/R^3}$ for $m=2$, where
$\omega$ is the frequency observed in the corotating frame of the star.
Here only the fundamental $r$-modes with $l'=m=2$ are considered since they are
most important for the $r$-mode instability of neutron stars
(see Lockitch \& Friedman 1999; Yoshida \& Lee 2000a, 2000b).
The crustal toroidal mode shown in Figure 1 is the fundamental mode with no radial nodes of
the eigenfunction and has the smallest oscillation frequency of
the mode of this kind for a given $l'$ at $\Omega=0$ (see McDermott et al. 1988).
Note that at $\bar{\Omega} \sim 0$ 
it is practically impossible to correctly calculate rotational modes 
because of their coupling with high overtone $g$-modes. 
In this figure, we have used the notation given by $_{l'}r^c_n$ and $_{l'}r^s_n$
for the $r$-modes, $_{l'}t_n$ for the crustal toroidal modes, and $_{l_0}i$ for the inertial 
modes, where $l'$ denotes the index of the spherical harmonic function
associated with the dominant toroidal component of the displacement vector, $n$
is the number of radial nodes of the eigenfunction, and $l_0$ is the
number introduced to classify the inertial modes (see, e.g., Yoshida \& Lee 2000a).
Note that the mode classification has been done at sufficiently small values of $\bar\Omega$.
Since there appear surface $r$-modes and core $r$-modes for the three-component
neutron star models, we have also introduced the superscripts $s$ and $c$ to distinguish
the two kinds of $r$-modes.
The oscillation energy of the core $r$-mode is predominantly
confined in the fluid core, and that of the surface $r$-mode is
confined in the surface ocean.
A typical eigenfunction of a surface $r$-mode is given in Lee \& Strohmayer (1996).
The modes in the figure are all odd modes (with $l'=|m|$ for the toroidal modes) 
and they are retrograde modes propagating in the opposite direction 
to that of stellar rotation.

Figure 1 clearly shows that the mode crossing between the $_2r_0^c$-mode and 
the $_2t_0$-mode, as predicted by equations (\ref{om-r}) to (\ref{om-cross}),
leads to an avoided crossing of their frequencies as functions of $\bar\Omega$.
At the avoided crossing, the modal properties of the two modes are exchanged.
This is understood by examining the eigenfunctions of the two modes before and after
the crossing, which are indicated by the filled circles in Figure 1, where the labels (a), (c),
and (d) attaching to the filled circles are corresponding to the panels (a), (c), and (d)
in Figure 2.
Note that the $_2t_0$-mode at $\bar\Omega=4\times10^{-3}$, which should carry the label (b), 
is not shown in Figure 1, because $\kappa\gg1$.
In Figure 2, the expansion coefficients $S_3$, $H_3$, and $iT_2$ of the two modes 
with $l'=m=2$ at $\bar\Omega=4\times10^{-3}$ in panels (a) and (b)
and at $\bar\Omega=4\times10^{-2}$ in panels (c) and (d)
are given as functions of the fractional radius $r/R$, where
the eigenfunctions are normalized as $iT_{l'}(r_{bc})=1$ with $r_{bc}$ being the 
bottom of the solid crust of the model.
When $\bar\Omega$ is sufficiently small, the $_2r_0^c$-mode in panel (a) and 
the $_2t_0$-mode in panel (b) have their typical eigenfunctions
and the influence of other modes on the eigenfunction is negligible.
Through the avoided crossing, however,
the modal properties of the two modes are exchanged along the frequency curves 
as $\bar\Omega$ increases.
As shown by Figure 2, the mode in panel (d), which is corresponding to the mode (d)
on the frequency curve of $_2t_0$ in Figure 1, has in the fluid core 
the eigenfunction characteristic of the $_2r_0^c$-mode.
On the other hand, the mode in panel (c), which corresponds to the mode (c) 
on the frequency curve of $_2r_0^c$, has in the crust the eigenfunction characteristic of
the $_2t_0$-mode.
However, this avoided crossing between the two modes is not sharp in the sense that
the eigenfunction of each of the two modes
suffers the contamination of the other mode after the avoided crossing.
For example, the mode (d), which may be regarded as a core $r$-mode, 
has also oscillation amplitudes in the solid crust.
This is in a sharp contrast with avoided crossings, for example,
between $_2r_0^s$ and $_2r_0^c$ and between $_5i$ and $_2r_0^c$, for which
the mode properties are almost completely exchanged through the crossings.
Because of the avoided crossing between the $_2r_0^c$-mode and the crustal toroidal modes,
the $_2r_0^c$-mode will lose its simple modal property of the eigenfrequency and 
the eigenfunction as a function of $\bar\Omega$.
We note that the fundamental $r$-mode with $l'=m$ of fluid stars has such a simple
modal property as a function of $\bar\Omega$
(e.g., Yoshida \& Lee 2000b; Karino et al. 2000).

Avoided crossing between the $_{l'}r_0^c$ and $_{l'}t_0$ modes with $l'=m$ 
may be common for neutron stars with a solid crust.
To show this, we calculate the $_{l'}r_0^c$ modes with $l'=m=2$
for $n=1$ polytropic neutron star models in which a solid crust is artificially embedded 
and isentropic structure is assumed. 
We assume the same physical parameters, such as the mass, radius, and the crust thickness,
as those of the model NS13T8.
For the shear modulus $\mu_0$, we assume that
the ratio $\mu_0/\rho^{4/3}$ is constant in the crust where $\rho$ is the mass density
(see e.g., McDermott et al. 1988). 
In Table 1, we tabulated the values of $\mu_0/\rho^{4/3}$ in the crust
and the frequencies $\omega_t(0)$ of the fundamental $_2t_0$ mode at $\bar\Omega=0$.
We note that $\omega_t(0)$ of the polytropic model
is approximately proportional to the value of $\sqrt{\mu_0/\rho^{4/3}}$.
Figure 3 gives the scaled frequencies $\kappa=\omega/\Omega$ of the $_2r_0^c$-mode 
as functions of $\bar\Omega$ for the polytropic models.
The bends of the frequency curves of the $r$-modes as $\Omega$ increases
are caused by the avoided crossing with the crustal toroidal modes $_2t_0$,
as found in Fig.1 for a realistic neutron star model.
As Figure 3 indicates the avoided crossing between the $_2r_0^c$-mode and 
the $_2t_0$-mode is a common phenomenon and occurs 
at around $\bar\Omega_{cross}$, which is determined by $\omega_t(0)$ for 
given $l'$ and $m$ (see equation (\ref{om-cross})).
The apparent difference in behavior of the $r$-mode frequencies as a function of 
$\Omega$ between the three polytropic models is therefore attributable to the difference 
in the location $\Omega_{cross}$ of the avoided crossing, that is, 
to the difference in the property $\mu_0/\rho^{4/3}$ of the solid crust.
Here, it is worth while to note that, as shown by Yoshida \& Lee (2000b),
the deviation of the model from isentropic structure
is not important for the fundamental $r$-modes with $l'=m$
and inertial modes, 
if the Schwarzschild discriminant $A$ of the nonisentropic model has 
sufficiently small absolute values as is expected in neutron stars. 
As a matter of fact, we can obtain 
similar results to those shown in Figure 1, even if we artificially set $A=0$ in the model 
NS13T8. 
%


\section{Dissipation Timescales of Core $r$-Modes}

For comparison with other studies on the $r$-mode instability,
let us derive an extrapolation formula for the damping timescale of the 
fundamental core $r$-mode with $l'=m=2$ by calculating the $r$-modes
at sufficiently small values of $\bar\Omega$, for which the mode coupling 
with the crustal toroidal mode is negligibly weak.
Then, using the same shear and bulk viscosity coefficients as those employed in
Yoshida \& Lee (2000a) and taking account of the viscous boundary layer (VBL) 
damping effects following the formulation given in Bildsten \& Ushomirsky (2000)
(see also Andersson et al (2000) for a correction),
we may approximately express the formula as (see e.g., Lindblom et al. 1998)
\begin{eqnarray}
\frac{1}{\tau} &=& \frac{1}{\tilde \tau_S} \left( \frac{10^8~\rm{K}}{T_c} \right)^2
+ \frac{1}{\tilde \tau_B} \left( \frac{T_c}{10^8~\rm{K}} \right)^6
\left( \frac{\Omega^2}{\pi G \bar{\rho}} \right)  \nonumber \\
&+& \frac{1}{\tilde \tau_{VBL}} \left( \frac{10^8~\rm{K}}{T_c} \right) 
\left( \frac{\Omega^2}{\pi G \bar{\rho}} \right)^{1/4}
+ \frac{1}{\tilde \tau_{J,2}} 
\left( \frac{\Omega^2}{\pi G \bar{\rho}} \right)^{3}  \, ,
\label{tau}
\end{eqnarray}
where $\bar{\rho}$ is the average density of the star, and $T_c$ is the central temperature.
The same viscosity coefficients are used both in the fluid and solid regions,
and no effects of superfluidity in the core and in the inner crust are considered.
Here, the first, second, third and 
fourth terms in the right-hand side of equation (\ref{tau}) are contributions 
from the shear viscosity, the bulk viscosity, the shear viscosity in VBL and the 
current quadrupole radiation, respectively. 
The factors containing the ratio $T_c/10^8~{\rm K}$ have been introduced
in equation (\ref{tau}) to extrapolate the damping rates
with respect to the central temperature $T_c$.

In Table 2, the various dissipation timescales $\tilde\tau$ 
of the $_2r_0^c$ modes with $l'=m=2$ are tabulated
for the model NS13T8 and a simple $n=1$ polytropic model with a static solid crust.
For the polytropic model, we employ $M=1.4 M_{\sun}$ and $R=12.5 \rm{km}$ and
assume that a solid crust forms in the density region of 
$\rho\le1.5\times 10^{14}~{\rm g/cm^3}$.
Note that no wave propagation is assumed 
in the static crust of the polytropic model employed in this section.
Bildsten \& Ushomirsky (2000) (see also Anderson et al. 2000, and Rieutord 2000 for more detailed 
treatment of VBL) used the similar polytropic
model with a static solid crust to estimate the VBL damping effects on the $r$-modes.
To compute the $r$-modes of the simple polytropic model, 
the method given in Yoshida \& Lee (2000a), in which effects of the centrifugal force and 
perturbations of the gravitational potential are taken into account, has been employed.   
As shown by Table 2, all the dissipative timescales, except that
due to the bulk viscosity, have similar values both for the simple polytropic model 
and for the model NS13T8.
We note that although the destabilization due to the current quadrupole radiation
for the model NS13T8 is by one order of magnitude stronger than that for
the polytropic model, the VBL damping timescales, which are most important
among the damping mechanisms of the $r$-modes,
are almost the same for the two models. 
We consider that the difference in $\tau_{J,2}$ between the two models is 
caused by the difference in the compactness $G M /(c^2 R)$ of the models, where 
$c$ denotes velocity of light.  
The huge difference in the damping timescales associated with the bulk viscosity 
between the two models
may come from the difference in the treatment of the modes in the crust.
Although we count the damping contributions in the crust for the model NS13T8, 
we do not count them for the polytropic model because we have assumed that the crust
is static and allows no oscillations.
In addition to this, 
the model NS13T8 in an evolutionary sequence of cooling neutron stars has
the crustal temperature that is higher by a factor 7 than the core temperature.
Since the bulk viscosity coefficient is proportional to $T^6$, the contribution
of the bulk viscosity in the crust is largely enhanced
for the damping of the $r$-modes.


\section{Discussion and Conclusion}

It is well known that retrograde oscillations 
whose frequency $\omega$ satisfies 
the condition $0 < \omega/\Omega < m$ are unstable to the gravitational radiation 
reaction (Friedman \& Schutz 1978). 
Using equation (\ref{om-t}), we can show that
the retrograde crustal toroidal modes satisfy 
this condition and become unstable to
the gravitational radiation reaction when
\begin{equation}
\Omega > \frac{l'\, (l'+1)}{m \, (l'^2+l'-1)} \, \omega_t(0) \ ,  
\label{om-stable}
\end{equation}
where $\omega_t(0)$ denotes the oscillation frequency of the mode at $\Omega=0$.
This instability can happen at slow rotation rates of the star if $\omega_t(0)$ is 
sufficiently small.
Note that all crustal toroidal modes shown in  Figure 1 are unstable to  
the gravitational radiation reaction. 
Our preliminary stability calculation, however, shows that the instability  
of the crustal toroidal mode is very weak and may not be important.

In this paper, we have investigated the properties of $r$-modes of
slowly rotating neutron stars with a solid crust. 
We have found that the mode property of the core $r_0$ mode with $l'=m=2$,
which is most important for the $r$-mode instability of neutron stars, 
is strongly affected by mode coupling with 
the crustal $_2t_0$ modes as $\Omega$ increases.
This means that we cannot assume a simple mode property of 
the core $r_0$ modes with $l'=m$ as a function of $\bar\Omega$.
To discuss the $r$-mode instability of rotating neutron stars
it has been common to derive an extrapolation formula 
for the damping timescale of the $r$-modes (e.g., Lindblom et al. 1998).
The extrapolation is usually carried out 
about the interior temperature $T$ and the rotation frequency $\Omega$.
The extrapolation with respect to the temperature may be justified if the stars have
an isothermal structure
characterized, for example, by the central temperature $T_c$.
On the other hand, since the mode properties of the $r$-mode are well 
known only in the limit of 
$\bar\Omega\rightarrow 0$, the damping timescales are calculated in this limit and are
extrapolated to rapidly rotating neutron stars with $\bar\Omega\sim1$.
This process of the extrapolation with respect to $\Omega$ is justified if
the mode frequency $\omega$ is well approximated by
$2m\Omega/l'(l'+1)$ and the eigenfunction has the dominant toroidal component, 
independent of $\Omega$.
We know that the above two assumptions are satisfied for the fundamental $r$-modes 
with $l'=m$ for fluid stars (Yoshida \& Lee 2000b; Yoshida et al. 2000; Karino et al. 2000).
In the case of neutron stars with a solid crust, however, the two assumptions are not 
satisfied except when $\bar\Omega\sim 0$ and no mode couplings are expected.
For example, when the $r$-mode is near the avoided crossing with the crustal $_2t_0$ mode, 
the eigenfunctions and eigenfrequencies of the $r$-mode behave peculiarly and 
change rapidly as functions of $\Omega$.
This makes it difficult to predict reliably
the evolution of a neutron star driven by the $r$-mode 
instability through the avoided crossing 
by simply extrapolating the damping timescales obtained at 
small $\Omega$.
In addition to this, the eigenfunctions of the $r$-modes are 
contaminated by the crustal torsional modes at large $\Omega$, and the 
extrapolated damping timescales are not 
necessarily accurate enough to give the useful estimations.
These difficulties mean that the extrapolation formula of 
the damping timescales of the $r$-modes,
derived in the limit of $\bar\Omega\rightarrow 0$,
is not applicable to the discussion on the $r$-mode instability of the rapidly rotating
neutron stars with a solid crust
unless the effects of the avoided crossings with and the contamination of 
the eigenfunctions by the crustal torsional modes are correctly taken into account. 
In this sense, discussions based on the extrapolation formula of the damping 
timescales are not fully justified for neutron stars with a solid crust.



%


\newpage

\begin{deluxetable}{cll}
\footnotesize
\tablecaption{Shear modulus $\mu_0$ and eigenfrequency $\omega_t(0)$ of 
the $_2 t_0$ mode for the polytropic model with a crust at $\Omega=0$.}
\tablewidth{300pt}
\tablehead{
 \colhead{Model}&\colhead{$\mu_0/\rho^{4/3}$}
                &\colhead{$\omega_t(0)/(G M/R^3)^{1/2}\ $}
} 
\startdata
model 1 & $ 1.0 \times 10^{11} $&$ 9.9 \times 10^{-3}$ \nl
model 2 & $ 1.0 \times 10^{12} $&$ 3.1 \times 10^{-2}$ \nl
model 3 & $ 1.0 \times 10^{13} $&$ 9.9 \times 10^{-2}$ \nl
\enddata
\label{model-p}
\end{deluxetable}

\begin{deluxetable}{cllll}
\footnotesize
\tablecaption{Dissipative timescales $\tilde\tau$ of core $r_0$ modes with $l'=m=2$ 
for neutron star models. }
\tablewidth{0pt}
\tablehead{
 \colhead{Model}  
 & \colhead{$\tilde \tau_B$(s)}  
 & \colhead{$\tilde \tau_S$(s)} & \colhead{$\tilde \tau_{VBL}$(s)}    
 & \colhead{$\tilde \tau_{J,\vert m\vert}$(s)}
} 
\startdata
NS13T8 &
$ 1.3 \times 10^{7} $&$ 1.6 \times 10^{5}$&$ 3.7\times 10 $&$ -2.2\times 10^{-1}  $ \nl
Polytrope with a static crust &
$ 1.7 \times 10^{17}$&$ 1.9 \times 10^{6}$&$ 8.9\times 10 $&$ -4.4\times 10^0$\nl
\enddata
\label{d-time}
\end{deluxetable}


\newpage

\begin{figure}
\epsscale{.7}
\plotone{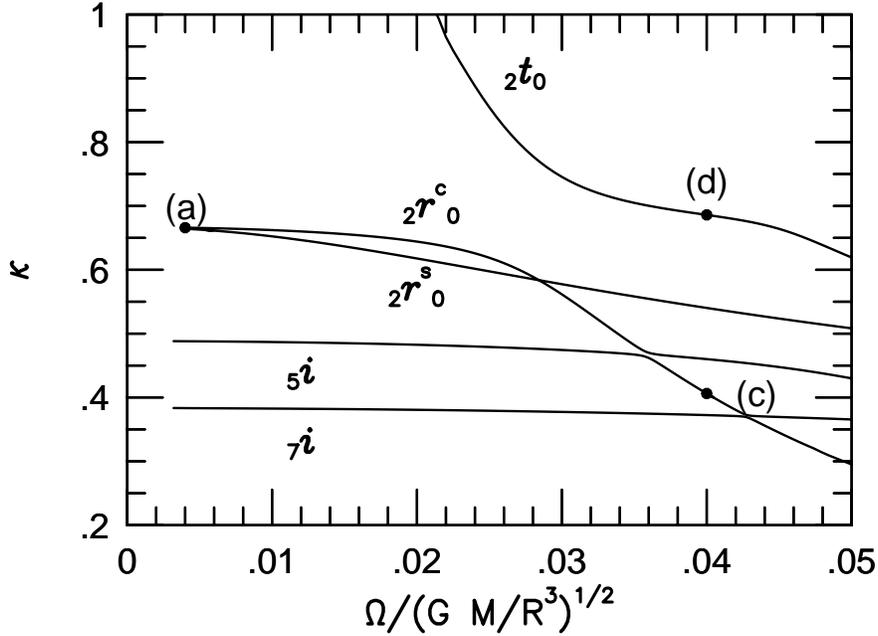}
\caption{Scaled frequencies $\kappa=\omega/\Omega$ of the $r$-modes, 
inertial modes and the crustal torsional (toroidal) mode of the model NS13T8
are plotted as functions of $\bar{\Omega}=\Omega/(GM/R^3)^{1/2}$ for $m=2$, where
the frequency curves are attached by the labels indicating 
their classification at small values of $\bar\Omega$ (see the text for 
the detail of the classification).
The modes depicted are retrograde odd modes.
The $r$-modes and the crustal toroidal mode have the dominant toroidal component associated
with $l'=m=2$ at small values of $\bar\Omega$.
Note that the eigenvalue of the $r$-modes is given by $\kappa=2 m/(l'(l'+1))$ in 
the limit of $\bar\Omega\rightarrow 0$. The filled circles indicate the modes
whose eigenfunctions are shown in Figure 2.}
\end{figure}

\newpage

\begin{figure} 
\epsscale{1} 
\plotone{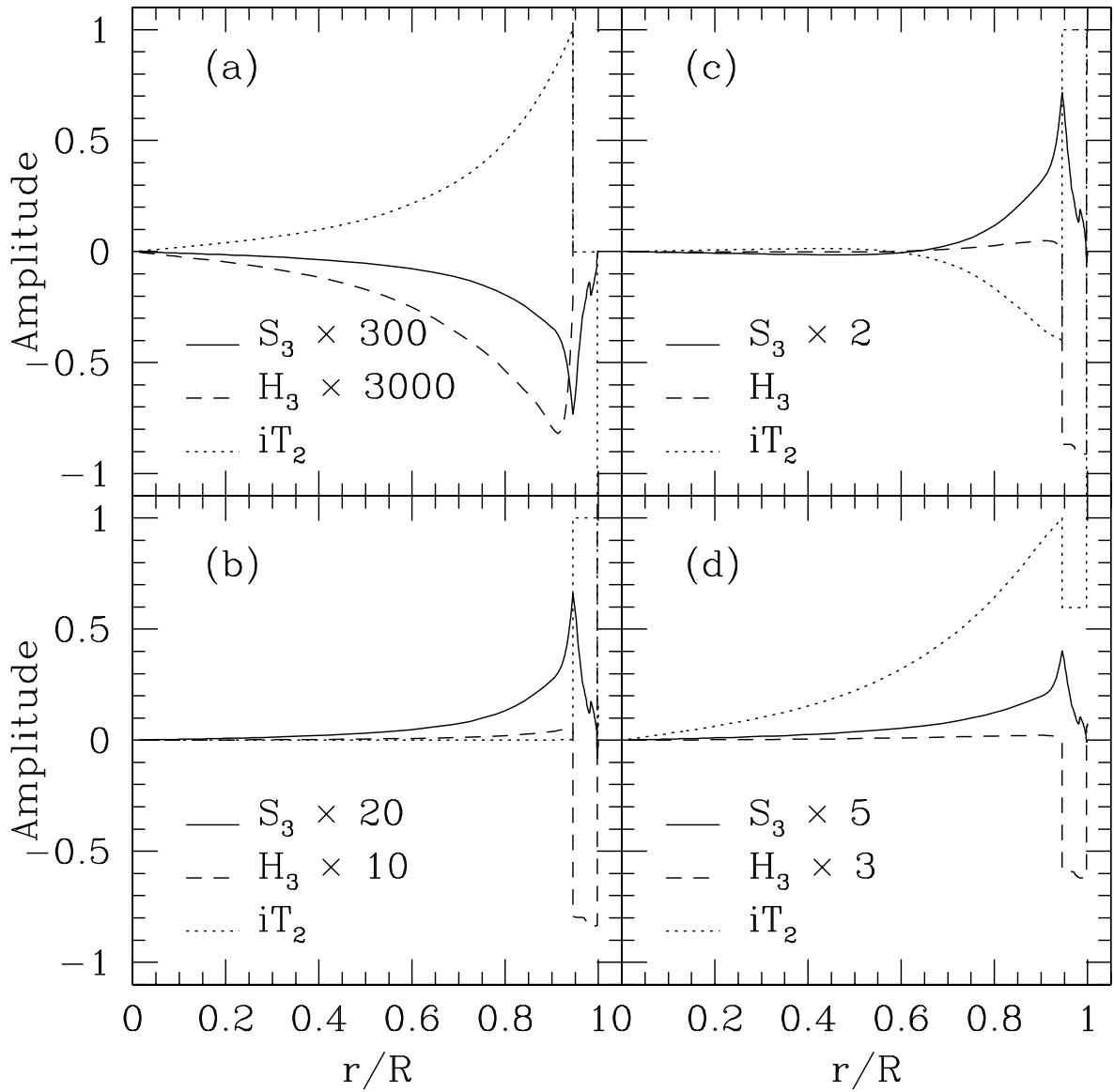} 
\caption{Expansion coefficients  
$S_3$ (solid curve), $H_3$ (dashed curve), and $iT_2$ (dotted curve) 
of the core $r$-modes and the crustal 
toroidal modes with $l'=m=2$ for the model NS13T8
are given as functions of $r/R$ at $\bar\Omega=4\times10^{-3}$ in panels (a) and (b), 
and at $\bar\Omega=4\times10^{-2}$ in panels (c) and (d), 
where the normalization of the eigenfunctions are 
given as $iT_2(r_{bc})=1$ with $r_{bc}$ being the bottom of the solid crust. 
The eigenfunctions shown in panels (a) and (b) are those of the $_2r_0^c$-mode and  
the $_2t_0$-mode, respectively.
The locations of the modes are shown in the $\kappa-\bar\Omega$ plane
by the filled circles in Figure 1.}
\end{figure}

\newpage

\begin{figure}
\epsscale{.7}
\plotone{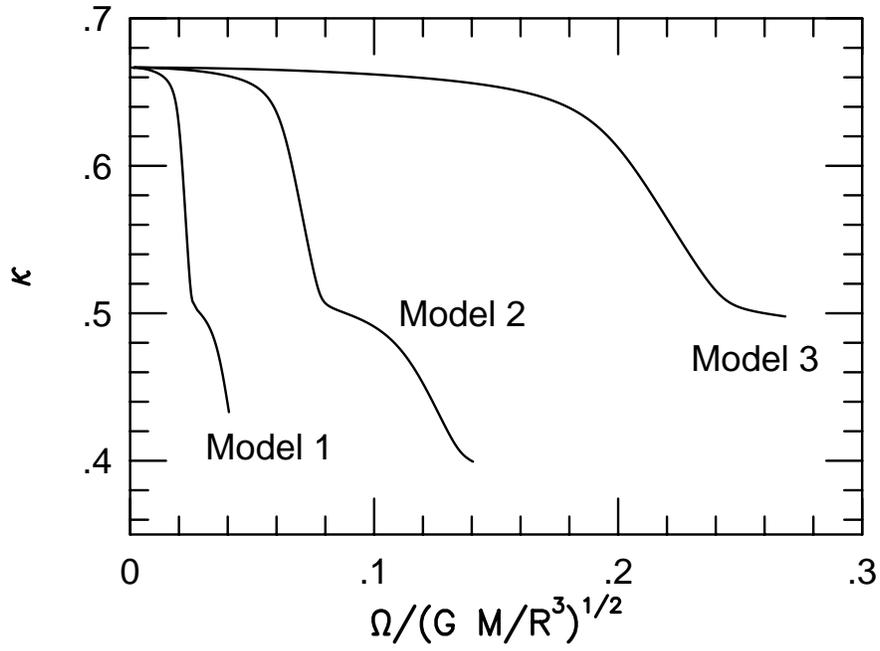}
\caption{Scaled frequencies $\kappa=\omega/\Omega$ for 
the fundamental $_2 r_0^c$ modes with $l'=m=2$ are plotted as functions of 
$\bar{\Omega}=\Omega/(GM/R^3)^{1/2}$ for the simple polytropic model with a solid crust. 
The three models have different values of $\mu_0/\rho^{4/3}$ that is assumed to be constant 
in the crust (see, Table 1).} 
\end{figure}

\end{document}